\documentclass[12pt,thmsa,a4paper,oneside,final,onecolumn,titlepage]{article}
\usepackage{sw20lart}

\input{tcilatex}
\begin{document}

\title{Bell's Theorem and Chemical Potential }
\author{A. Shafiee\thanks{%
E-mail: shafiee@theory.ipm.ac.ir} $^{\text{(1,3)}}$ \ \ \ M. Golshani$^{%
\text{(2,3)}}$\quad M. G. Mahjani$^{\text{(4)}}\medskip $ \\
{\small \ }$\stackrel{1)}{}$ {\small Department of Chemistry, Kashan
University,}\\
{\small \ Kashan, 87317-51167, Iran.}\\
{\small \ }$\stackrel{2)}{}$ {\small Department of Physics, Sharif
University of Technology,}\\
{\small \ P.O.Box 11365-9161, Tehran, Iran.}\\
{\small \ }$\stackrel{3)}{}$ {\small Institute for Studies in Theoretical
Physics \& Mathematics,}\\
{\small \ P.O.Box 19395-5531, Tehran, Iran.}\\
{\small \ }$\stackrel{4)}{}$ {\small Department of Chemistry, K.N.Toosi
University of Technology, }\\
{\small \ P.O.Box 1587-4416, Tehran, Iran.}}
\maketitle

\begin{abstract}
Chemical potential is a property which involves the effect of interaction
between the components of a system, and it results from the whole system. In
this paper, we argue that for two particles which have interacted via their
spins and are now spatially separated ,the so-called Bell's locality
condition implies that the chemical potential of each particle is an
individual property. Here is a point where quantum statistical mechanics and
the local hidden variable theories are in conflict. Based on two distinct
concepts of chemical potential, the two theories predict two different
patterns for the energy levels of a system of two entangled particles. In
this manner, we show how one can distinguish the non-separable features of a
two-particle system.
\end{abstract}

\section{\protect\smallskip Introduction}

Bell's theorem [1, 2] has a distinguished place in the contemporary research
on the foundations of physics. In general terms, this theorem concerns two
spin $\frac{1}{2}$ particles which were once together and now have a
space-like separation. According to this theorem, one cannot, \textit{using a%
} \textit{certain definition of locality}, construct a hidden variable
theory that can reproduce all the predictions of quantum mechanics (QM). By
a \textit{certain definition of locality}, we mean ''Bell's locality
condition'' which in a local stochastic hidden variable (LSHV) theory is
equivalent to the statistical independence of the values of the spin
components of the two particles. While, people agree that according to QM
Bell's locality condition cannot hold at a sub-quantum level, there is no
unanimous agreement on the scope of such non-locality [3]. Does it mean the
existence of superluminal communication between the two particles or it
implies that two particles that have once interacted can never be considered
to be independent of each other (or being separable), even though there is
no exchange of information between them. Here, we are confronted with two
quite distinct interpretations of Bell's locality condition which are
usually referred to as ``locality'' and ``separability'' in the literature
[4]. In a multi-component system, the locality assumption means that for a
given component, the value of an observable does not depend on the
measurements which are performed simultaneously on any other spatially
separated component. This is Einstein's view of locality. In his view,
however, the separability criterion implies that each component in a
space-time region has its own intrinsic physical state and that the entire
and complete physical state of a multi-component system is specified once
one has determined the intrinsic state of each component [5]. In Bell's
theorem, the two notions seems to be indistinguishable [6]: The spin
correlations are the properties of the whole system, i.e., the correlations
result from the entire singlet state, while the empirical verification of
them would be possible once the spin measurements are performed
simultaneously on two particles which are at space-like separation.

There have been many attempts to give an interpretation of the context of
Bell's theorem [7-9]. Here, we try to present a new perspective of this
matter by explaining how it is possible to identify the non-separable trait
of Bell's locality condition in a proposed experiment. Our argument shows
that the distinction between the notions of non-locality and
non-separability is actually realizable. Furthermore, our work opens a new
outlook of Bell's theorem in a broader sense in which the effective
spin-spin interaction between any two particles with the same spin can be
considered as a special factor for describing the non-separable nature of
the composite systems. The role of such interactions in understanding and
formulation of the chemistry of solutions and mixtures is an open problem.

In our paper, we consider a system of two spin $\frac{1}{2}$ particles which
have been interacted via their spins in the past and then are spatially
separated from each other. For such a system, we \textit{do not} consider
the spin correlations, i.e., we suppose \textit{no} spin measurements are
made on each individual particle. We introduce the idea of using the
chemical potential as a classical property, instead of using spin
correlations which have a quantum mechanical origin. Then, we carry out some
simple quantum statistical mechanics (QSM) calculations for a singlet state
to show that Bell's locality condition is equivalent to the assumption that
those properties of each particle which result from the whole system can
effectively be taken as individual properties. A physical consequence of
this is that the energy pattern of the system can be obtained from the
energy states of the individual particles. This is an implication of the
separability criterion which is in conflict with what we get from QSM.

In section 2, we consider a method for producing pairs of entangled spin $%
\frac{1}{2}$ particles in a singlet state and we review the QSM calculations
for an effective spin-spin Hamiltonian. In section 3, we calculate the
chemical potential of each particle in the quantum limit. We shall argue
that Bell's locality condition is equivalent to the separability criterion
when the whole system is taken into account. The non-separable character of
the system is recognizable through a unique energy pattern which, in turn,
can be identified by an appropriate spectroscopic measurement. Finally, in
the last section, we review the significance of our analysis.

\section{QSM Calculations for an Effective Spin-Spin Interaction}

Suppose we produce an ensemble of systems, each system containing a pair of
spin $\frac{1}{2}$ particles (e.g., two electrons). We send each pair of two
spin half particles towards an entangler which is assumed to be a device in
which the two particles interact via their spins and thereby the entangled
states are generated. After the interaction, we let the particles recede
from each other. We assume that the temperature of ensemble is very low and
the other experimental conditions (if relevant) are so adjusted that the two
emerging particles are in a singlet state.

We represent the spin-spin interaction of a pair of particles of our
ensemble by the Hamiltonian

\begin{equation}
H_{int}=\alpha \ \overrightarrow{\sigma }^{(1)}.\overrightarrow{\sigma }%
^{(2)}
\end{equation}
where $\overrightarrow{\sigma }^{(i)}$ represent Pauli spin matrices of the $%
i$th particle ($i=1,\ 2$) and $\alpha $ is the exchange coupling coefficient
between the two spins $\overrightarrow{\sigma }^{(1)}$and $\overrightarrow{%
\sigma }^{(2)}$and here we assume to be a positive constant for a specific
pair of spin $\frac{1}{2}$ particles. Here, by $\overrightarrow{\sigma }%
^{(1)}.\overrightarrow{\sigma }^{(2)}$ we mean $\sigma _{x}^{(1)}\sigma
_{x}^{(2)}+\sigma _{y}^{(1)}\sigma _{y}^{(2)}+\sigma _{z}^{(1)}\sigma
_{z}^{(2)}$.

A particular realization of producing the singlet states which is relevant
to our discussion has been recently provided by using the coupled quantum
dot systems [10, 11]. Quantum dots are small semiconductor structures which
can host a single electron in a three dimensional confined region [12]. If
two nearby single electrons in each quantum dot are being weakly coupled
(e.g., by tunneling between the dots), a double-dot system will be produced
which is the result of a combined action of the Coulomb interaction and the
Pauli exclusion principle. It has been shown that at low temperatures
(typically about $0.2$ K) and in absence of magnetic fields, the ground
state of a double-dot system is a spin singlet while the exited state is a
spin triplet. Each pair of the entangled electrons can be injected into two
distinct leads (one electron in each lead) and, then, the outgoing electrons
are separated [13].

Considering the dynamics of the spins of two electrons which are confined in
a double-dot system, the real Hamiltonian of the system can be replaced by
the effective Heisenberg Hamiltonian (1), where $\alpha $ is equal to $\frac{%
1}{4}$ of the difference between the triplet and the singlet states. The
order of magnitude of $\alpha $ is about $0.05$ meV, which is a typical
value for the exchange energy between two electrons in a double-dot system
[11].

In QSM, the density matrix of a coupled system, which is described by the
Hamiltonian (1), can be written as

\begin{equation}
\rho =\frac{e^{-\alpha \beta \overrightarrow{\sigma }^{(1)}.\overrightarrow{%
\sigma }^{(2)}}}{z}
\end{equation}
where $\beta =\dfrac{1}{kT}$, $T$ is temperature, $k$ is Boltzmann constant
and $z=Tr(e^{-\beta H_{int}})$ denotes the partition function of the
two-particle system.

Using the properties of Pauli matrices, one can show that

\begin{equation}
e^{-\alpha \beta \sigma _{j}^{(1)}\sigma _{j}^{(2)}}=\cosh (\alpha \beta
)-\sigma _{j}^{(1)}\sigma _{j}^{(2)}\sinh (\alpha \beta )
\end{equation}
where $j=x,y,z$; $\cosh (\alpha \beta )=\dfrac{e^{\alpha \beta }+e^{-\alpha
\beta }}{2}$ and $\sinh (\alpha \beta )=\dfrac{e^{\alpha \beta }-e^{-\alpha
\beta }}{2}$.

Inserting (3) in (2), we obtain

\begin{equation}
e^{-\alpha \beta \overrightarrow{\sigma }^{(1)}.\overrightarrow{\sigma }%
^{(2)}}=\stackunder{j=x,y,z}{\dprod }\left[ \cosh (\alpha \beta )-\sigma
_{j}^{(1)}\sigma _{j}^{(2)}\sinh (\alpha \beta )\right]
\end{equation}

Using the properties of Pauli matrices again, (4) reduces to

\begin{equation}
e^{-\alpha \beta \overrightarrow{\sigma }^{(1)}.\overrightarrow{\sigma }%
^{(2)}}=\dfrac{1}{4}\left( e^{3\alpha \beta }+3e^{-\alpha \beta }\right)
\left( 1-\overrightarrow{\sigma }^{(1)}.\overrightarrow{\sigma }^{(2)}\
S_{\alpha \beta }\right)
\end{equation}
where $S_{\alpha \beta }=\dfrac{e^{2\alpha \beta }-e^{-2\alpha \beta }}{%
e^{2\alpha \beta }+3e^{-2\alpha \beta }}.$ The eigenstates of $%
\overrightarrow{\sigma }^{(1)}.\overrightarrow{\sigma }^{(2)}$ consist of $%
|\Phi _{1}\rangle =|++\rangle ,|\Phi _{2}\rangle =|--\rangle $ and $|\Phi
_{3}\rangle =\dfrac{1}{\sqrt{2}}\left[ |+-\rangle +|-+\rangle \right] $ with
the eigenvalue +1 and $|\Phi _{4}\rangle =\dfrac{1}{\sqrt{2}}\left[
|+-\rangle -|-+\rangle \right] $ with the eigenvalue -3. Here, $|++\rangle $
indicates that the values of the z- components of both $\sigma _{z}^{(1)}$
and $\sigma _{z}^{(2)}$ are $+1$, and a similar definition holds for the
other cases.

Using a complete set of these eigenstates, we can calculate the partition
function $z$:

\begin{equation}
z=Tr\left( e^{-\alpha \beta \overrightarrow{\sigma }^{(1)}.\overrightarrow{%
\sigma }^{(2)}}\right) =e^{3\alpha \beta }+3e^{-\alpha \beta }
\end{equation}

Thus, (2) is reduced to

\begin{equation}
\rho =\dfrac{1}{4}\left( 1-\overrightarrow{\sigma }^{(1)}.\overrightarrow{%
\sigma }^{(2)}\ S_{\alpha \beta }\right)
\end{equation}

At very low temperatures ($T\rightarrow 0$ or $\beta \rightarrow \infty $),
we have $\stackunder{\beta \rightarrow \infty }{\lim }S_{\alpha \beta }=1$.
This limit is known as quantum limit. According to QSM and in the quantum
limit, we have

\begin{equation}
\rho _{QM}=\dfrac{1}{4}\left( 1-\overrightarrow{\sigma }^{(1)}.%
\overrightarrow{\sigma }^{(2)}\right)
\end{equation}

The states $|\Phi _{1}\rangle $, $|\Phi _{2}\rangle $ and $|\Phi _{3}\rangle 
$ are three eigenstates of $\rho _{QM}$ with the eigenvalue zero and $|\Phi
_{4}\rangle $ is the singlet state with the eigenvalue +1. Thus $\rho _{QM}$
can be written as $\rho _{QM}=|\Psi _{0}\rangle \langle \Psi _{0}|,$ where $%
|\Psi _{0}\rangle =|\Phi _{4}\rangle $, i.e., the singlet state. This can
also be shown by writing $|+\rangle \langle +|=\dfrac{1}{2}(1+\sigma _{z}),$ 
$|-\rangle \langle -|=\dfrac{1}{2}(1-\sigma _{z}),$ $|+\rangle \langle -|=%
\dfrac{1}{2}(\sigma _{x}+i\sigma _{y})$ and $|-\rangle \langle +|=\dfrac{1}{2%
}(\sigma _{x}-i\sigma _{y})$ and substituting these in the $|\Psi
_{0}\rangle \langle \Psi _{0}|$ expression.

The relation (8) shows that for a pair of spin $\frac{1}{2}$ particles which
have interacted through the Hamiltonian (1), the spin state of the system in
the quantum limit is described by a singlet state. When these two spin $%
\frac{1}{2}$ particles interact, we get two energy levels: One having a
lower energy $-3\alpha $ belonging to the singlet state; the other one
having a higher energy $+\alpha $, belonging to the triplet state. At very
low temperatures, the lowest occupied energy state is the one having the
energy $-3\alpha $ and so the quantum state of the system is a pure singlet
state. When no perturbation is introduced, the system is going to remain in
the singlet state. But, if the temperature is raised the pure singlet state
is lost. Similarly, once we measure a spin component of a particle, there is
going to be an interaction with an external field (e.g., a magnetic field in
the case of Stern-Gerlach apparatus). Then, the energy pattern of the system
is changed and a new pattern will be formed which is composed of the
single-particle energy states.

\section{QSM Versus LSHV Theories}

Consider a canonical ensemble consisting of $N$ identical distinguishable
coupled systems, e.g. $N$ double quantum dot systems in a semiconductor
heterostructure [12] which are distinguishable because of their locations.
The partition function of the canonical ensemble is $Z=z^{N},$ where $z$ is
the partition function of the two-particle system and is obtained from (6).
If we denote the number of particles 1 and 2 in the ensemble by $N_{1}$ and $%
N_{2}$, respectively, then $N=\dfrac{N_{1}+N_{2}}{2}$. For this ensemble,
the chemical potential of each particle resulting from the spin interaction
of each pair (which is a coupling process), under the condition of a
definite temperature, is equal to

\begin{equation}
\mu _{i}=-\dfrac{1}{\beta }(\dfrac{\partial \ln Z}{\partial N_{i}})=-\dfrac{1%
}{2\beta }\ln z
\end{equation}
where $i=1,2.$ Using (6) and (9), we get

\begin{equation}
\mu _{1}=\mu _{2}=-\dfrac{3}{2}\alpha -\dfrac{1}{2\beta }\ln (1+3e^{-4\alpha
\beta })
\end{equation}

In the limit $\beta \rightarrow \infty ,$ the chemical potential reduces to $%
-\dfrac{3}{2}\alpha .$ Chemical potential can be defined for any species of
particles in the pair, but \textit{it is not an individual property}. From
(9), it is clear that the chemical potential of each particle can be
calculated from the partition function of the two-particle system, and here
one cannot reduce $z$ to the single particle case. In other words, the
chemical potential for each particle is a property which involves the effect
of interaction and results from the whole system. From (10), it is clear
that the effect of interaction on the chemical potential enters through the
parameter $\alpha $ and in the quantum limit, it approaches the value $-%
\dfrac{3}{2}\alpha .$ On the other hand, in the quantum limit, the canonical
ensemble reduces to a pure ensemble in which only the ground state is
occupied. In this case, we have the maximum amount of order in the system
and the entropy of each system, which is $S=-k\ Tr(\rho _{QM}\ln \rho _{QM})$%
, is equal to zero. Thus, due to the existence of thermodynamic equilibrium
in the system, the chemical potential of the whole system, which is in the
ground state, is equal to the sum of single-particle chemical potentials in
a coupling process, and we have $\mu _{1}+\mu _{2}=-3\alpha .$

Now, Suppose that we have a sub-quantum level describable by some hidden
variables, in which the principle of separability is honored. After the
spin-spin interaction of the two-particle system and when the two particles
are spatially separated, we assume that the joint probability for the spin
of particle 1 along $\widehat{a}$ ($\sigma _{a}^{(1)}$) being $r$ ($r=\pm 1$%
) and the spin of particle 2 along $\widehat{b}$ ($\sigma _{b}^{(2)}$) being 
$q$ ($q=\pm 1$), is equal to

\begin{equation}
p_{r,q}^{(1,2)}(\widehat{a},\widehat{b},\lambda _{r}^{(1)},\lambda
_{q}^{(2)})=p_{r}^{(1)}(\widehat{a},\lambda _{r}^{(1)})\ p_{q}^{(2)}(%
\widehat{b},\lambda _{q}^{(2)})
\end{equation}
where $p_{r}^{(1)}(\widehat{a},\lambda _{r}^{(1)})$ and $p_{q}^{(2)}(%
\widehat{b},\lambda _{q}^{(2)})$ are individual probabilities for particles
1 and 2, respectively. Here, we are assuming that the spin state
corresponding to $\sigma _{a}^{(1)}=+1$ or $-1$ ($\sigma _{b}^{(2)}=+1$ or $%
-1$) for particle 1 (2) could be made to correspond to \textit{different}
collection of hidden variables $\lambda _{+}^{(1)}$ or $\lambda _{-}^{(1)}$ (%
$\lambda _{+}^{(2)}$ or $\lambda _{-}^{(2)}$) respectively, although in
Bell's theorem, they are usually characterized by a unique set of hidden
variables $\lambda $. Thus, with a broader attitude, we consider the
relation (11) as a criterion for the so-called ``Bell's locality condition''
in a LSHV theory. In a canonical ensemble, the individual probabilities are
equal to

\begin{equation}
p_{r}^{(1)}(\widehat{a},\lambda _{r}^{(1)})=\dfrac{e^{-\beta \epsilon
_{r}^{(1)}(\widehat{a},\lambda _{r}^{(1)})}}{z^{(1)}}
\end{equation}
and

\begin{equation}
p_{q}^{(2)}(\widehat{b},\lambda _{q}^{(2)})=\dfrac{e^{-\beta \epsilon
_{q}^{(2)}(\widehat{b},\lambda _{q}^{(2)})}}{z^{(2)}}
\end{equation}
In (12), $\epsilon _{r}^{(1)}(\widehat{a},\lambda _{r}^{(1)})$ refers to two
energy levels $\epsilon _{+}^{(1)}(\widehat{a},\lambda _{+}^{(1)})$
(corresponding to $\sigma _{a}^{(1)}=+1$) and $\epsilon _{-}^{(1)}(\widehat{a%
},\lambda _{-}^{(1)})$ (corresponding to $\sigma _{a}^{(1)}=-1$). It is
assumed that these levels \textit{split} due to the spin-spin interaction of
the pair. Also, $z^{(1)}$ is the partition function of the particle (1) and
is equal to $z^{(1)}=e^{-\beta \epsilon _{+}^{(1)}}+e^{-\beta \epsilon
_{-}^{(1)}}.$ The same description holds for (13). In the canonical
ensemble, the chemical potential of the first particle is equal to

\begin{eqnarray}
\mu _{1} &=&-\dfrac{1}{\beta }\ln z^{(1)}  \nonumber \\
&=&\epsilon _{+}^{(1)}-\dfrac{1}{\beta }\ln \left[ 1+e^{-\beta (\epsilon
_{-}^{(1)}-\epsilon _{+}^{(1)})}\right]
\end{eqnarray}
In the quantum limit, $\beta \rightarrow \infty $, depending on whether $%
\epsilon _{+}^{(1)}$ is lower or $\epsilon _{-}^{(1)},$ $\mu _{1}$ becomes
equal to one of them. In this case, $\mu _{1}$ is numerically equal to the
energy of the occupied ground state. According to (10), in the limit $\beta
\rightarrow \infty $, this value must be equal to $-\dfrac{3}{2}\alpha .$
Thus, the spin energy level of the first particle, after the interaction
with the second particle, splits into two levels: $-\dfrac{3}{2}\alpha $ and 
$+\dfrac{3}{2}\alpha $. But, these are the corresponding hidden variables
that determine which of these two levels belongs to $\epsilon _{+}^{(1)}$
and which one belongs to $\epsilon _{-}^{(1)}$. The same result holds for
the second particle. Thus, we have:

\begin{equation}
z^{(1)}=z^{(2)}=e^{-\frac{3}{2}\alpha \beta }+e^{+\frac{3}{2}\alpha \beta }
\end{equation}

In the relation (15) and consequently the relations (12) and (13), the
effect of interaction is introduced through the parameter $\alpha $. It is
also important to notice that the equation obtained for $z^{(1)}$ and $%
z^{(2)}$ in (15) is \textit{independent} of what was assumed about lambda in
a LSHV theory, but it is based on the crucial expressions of chemical
potential in relations (10) and (14).

Here, we are dealing with \textit{two different concepts} for $\mu _{1}$ and 
$\mu _{2}$. In the sub-quantum level, Bell's locality condition requires
that the chemical potential of each particle to be an \textit{individual
property} and that in quantum limit, it represents the lowest occupied
energy state for the particle. On the other hand, QSM calculations show that
for a system of two entangled particles there are \textit{no} individual
energy levels in the limit $\beta \rightarrow \infty $, and that $\mu _{1}$
and $\mu _{2}$ are \textit{not} representing the ground state energy of
single particles.

According to Bell's locality condition, the partition function of a
canonical ensemble is equal to

\begin{equation}
z=z^{(1)}z^{(2)}=e^{-3\alpha \beta }+e^{+3\alpha \beta }+2
\end{equation}

Thus, we are dealing with two different patterns for the energy levels.
According to (16), the system has three energy levels with the following
characteristics:

\begin{quote}
--- The ground state with an energy of $-3\alpha $, where each particle has
an energy of $-\dfrac{3}{2}\alpha $.

--- The first exited level, involving two states of zero energy, where the
first particle has an energy of $-\dfrac{3}{2}\alpha $ and the second
particle has an energy of $+\dfrac{3}{2}\alpha $ and vice versa.

--- The second excited state with an energy equal to $+3\alpha $,
corresponding to the case where each particle has an energy of $+\dfrac{3}{2}%
\alpha $. (Fig. 1-a)
\end{quote}

The aforementioned description for the energy pattern of the system, which
results from the different possible combinations of the single-particle
energy levels, is an explicit result of the separability condition. On the
other hand, according to (6), the system has two energy levels: The ground
state with an energy of $-3\alpha $, and the first excited level including
three states of an energy $+\alpha $. (Fig. 1-b) This pattern cannot be
deduced from the sums of individiual energy states.

In a LSHV theory, Bell's locality condition implies the first pattern which
in turn is a consequence of the separability criterion, but, QSM admits the
second one. There is an obvious difference in the physics of the problem.
Now, suppose, e.g., that the two-particle system is in the ground state
(i.e., in the quantum level). Furthermore, suppose that a radiation of an
energy $+3\alpha $ is incident on the system. In the first model, there is
always a probability for the system to be excited to the zero-energy level
(i.e., one of the particles remains in the level $-\dfrac{3}{2}\alpha $ and
the other one goes to the level $+\dfrac{3}{2}\alpha $), and then, returns
to the initial state by the emission of a photon with the energy $+3\alpha $%
. But this process does not take place in the second model, i.e., there
would be no change in the system. The spectroscopic study of the coupled
quantum dot systems has been cited in ref. [11].

\section{Conclusion}

We are concerned here with the concept of chemical potential. In a
multi-component system, the chemical potential attributed to each component
is generally a property which results from the whole system, i.e., it cannot
be interpreted to represent an individual property of a specific component
of the system. But, Bell's locality condition implies that chemical
potential can be attributed to each component of the system as an individual
property. Here, we use chemical potential as a characteristic feature of the
composite systems to demonstrate the (non)separability notion.

Furthermore, after the pioneering work of Guggenheim for electrochemical
processes in 1929 [14], it became clear that chemical potential is a
quantity in which the interaction with the field can be also included. Then,
the formulation of chemical potential was extended to include the potentials
due to the interacting fields.

For two spin $\frac{1}{2}$ particles that interact via their spins at low
temperatures and then recede from each other, the calculation of the
chemical potential of a particle in a canonical ensemble shows that once the
system is described by a singlet state, the effect of the interaction
remains in the particles, even if they are spatially separated. Here, there
is a common characteristic for the two particles: they are constrained to a
unique energy pattern. This is not due to the exchange of information
between the particles; rather, it is due to their past interaction which
implies the non-separable nature of a two-particle system. In the singlet
state, the energy pattern of the system is not reducible. It is a property
of the whole system and in the limit of low temperatures, the values of the
chemical potential of single particles coincides with none of the individual
energy levels. On the other hand, Bell's locality condition requires that at
the sub-quantum level, when no spin measurements are performed on each
particle, the energy pattern of the whole system can be obtained from the
combination of the energy levels of the individual particles. This is a
consequence of the separability assumption.

Thus, Bell's theorem is not confined to the evaluation and the comparison of
spin correlations in the quantum and sub-quantum levels; rather, it can be
used to distinguish the non-separable nature of composite systems which in
turn sheds new light on the physics and the chemistry of complex systems,
like solutions and mixtures. Chemical potential is a key property in this
context.\bigskip

\textbf{Acknowledgment\bigskip }

We would like to thank Prof. Hashem Rafii-Tabar for stimulating discussion
about the subject of the paper.

\end{document}